\documentclass[10pt,tightenlines,
nofootinbib,
prd,
twocolumn,
showpacs,preprintnumbers]{revtex4}

\newcommand{\beq}{\begin{equation}} 
\newcommand{\eeq}{\end{equation}} 
\newcommand{\bea}{\begin{eqnarray}} 
\newcommand{\eea}{\end{eqnarray}} 
\newcommand{\nn}{\nonumber} 
\newcommand{\junk}[1]{} 
\def\<{\langle} 
\def\>{\rangle} 
\def\d{\partial} 
\def\+{\dagger} 

\def\Tc{$T_c~$}
\def\neel{\vec{m}}
\def\gtil{\tilde{g}}
\def\gcrit{\tilde{g}_{\mathrm{crit}}}
\def\rvec{\vec{r}}

\def\U1{$U(1)$}

\def\UY{$U(1)_{Y}~$}

\def\mueff{\mu_{\mathrm{eff}}}

\def\meffk0{m_{\mathrm{eff}}^{0^{2}}}
\def\kzero{K^0}
\def\kplus{K^+}
\def\vpi{v}
\def\fpi{f_{\pi}}
\def\thetak{\theta_{K^0}}

\def\thetacrit{\theta_{\mathrm{crit}}}
\def\rtil{\tilde{r}}

\def\trace{\mathrm{Tr}}
\def\diag{\mathrm{diag}}

\begin{document} 
 
\title{Vortons in the $SO(5)$ model of high temperature superconductivity} 
\author{Kirk~B.~W.~Buckley and Ariel~R.~Zhitnitsky} 
\affiliation{Department of Physics and Astronomy, \\
University of British Columbia, \\
Vancouver, BC V6T 1Z1, Canada  }
\date{\today}
\begin{abstract} 
It has been shown that superconducting vortices with antiferromagnetic cores
arise within Zhang's $SO(5)$ model of high temperature 
supercondictivity. Similar phenomena where the symmetry is 
not restored in the core of the vortex was discussed by Witten in the case 
of cosmic strings. It was also suggested that such strings can form
stable vortons, which are closed loops of such vortices.
Motivated by this analogy, in following we
will show that loops of such vortices in the $SO(5)$ model of high $T_c$ 
superconductivity can exist as classically stable 
objects, stabilized by the presence of conserved charges trapped on the 
vortex core. These objects carry angular momentum which
counteracts the effect of the string tension
that causes the loops to shrink. 
The existence of such quasiparticles, which  are called vortons, 
could be interesting for the physics of high temperature superconductors.
We also  speculate that the phase transition between superconducting 
and antiferromagnetic phases at zero external magnetic field
when the doping parameter changes is 
associated with vortons.
\end{abstract} 
 
\maketitle 

\section{Introduction}

In the pursuit of a unified theory of high temperature superconductivity
and antiferromagnetism, 
Zhang proposed the $SO(5)$ theory of antiferromagnetism (AF) and d-wave 
superconductivity (dSC) in the high \Tc cuprates \cite{zhang}. 
The order parameter for 
antiferromagnetism is the Neel vector $\neel$ which is a vector 
under the action of the $SO(3)$, the group of 3-d spatial rotations. On the
superconducting side, the relevant order parameter is the complex
superconducting order parameter $\psi$, which describes the gap in the 
electron spectrum. The effective Lagrangian for $\psi$ is invariant
under the group $U(1)$. The big step that Zhang originally proposed 
is that the two symmetry groups can be combined within a larger symmetry 
group, namely $SO(5)$. This means that the three component vector $\neel$ and
the complex order parameter $\psi$ can be combined to form a ``superspin''
vector $\vec{n}=(\psi_1,m_1,m_2,m_3,\psi_2)$
which transforms under the group $SO(5)$. The presence of doping in
the cuprates actually breaks this symmetry down to $SO(3)\times U(1)$. 
At low doping, the AF phase is favored, corresponding to nonzero expectation
value for $|\neel|$, ($\<|\psi|\>=0$ and $\<|\neel|\> \neq 0$). 
As the doping is increased eventually the dSC phase
becomes energetically favorable with $\<|\psi|\> \neq 0$ and $\<|\neel|\> = 0$.
As Zhang originally discussed in \cite{zhang}, the region of intermediate
doping (near the AF-dSC phase boundary) should be characterized by 
conventional superconducting vortices, but possessing antiferromagnetic
cores. This suggestion was verified by various groups who looked
for numerical solutions of the classical equations of motion for 
different parameters \cite{arovas,berlinsky,cline}. Furthermore,
there has been recent experimental evidence that suggests
this theoretical picture may be correct 
\cite{vaknin,lake,dai,mitrovic,miller}. 

Our interest in this  topic arises from recent work 
\cite{fzstrings,krstrings,superk,qcdvorton} within 
a completely different context, namely the theory of the strong
interaction, QCD, at high baryon density \cite{colorsc,CFL}
(see \cite{colorscrev} for a good 
review of high density QCD and a long list of references). 
In \cite{superk} we  have 
shown that similar vortices with nontrivial core structure are present 
within QCD at high baryon density for physical values of the parameters. 
In this case, the symmetry breaking parameter responsible for the anisotropy
is the difference between the  up and down  quark masses. 
The effective Lagrangian which  describes the $SO(5)$ theory 
of high-\Tc superconductivity is very similar to the one used 
in \cite{superk}, aside from numerical constants of course. This will 
prove to be useful in making analogies between condensed matter physics 
and particle physics throughout the course of this work. 
We will make use
of the results given in \cite{superk, qcdvorton} throughout this paper. 

In the present paper we will show that is it possible to have loops of 
dSC vortices with AF cores that are classically 
stable objects. The source of this stability is the presence
of conserved charges trapped on the vortex core, leading to  nonzero angular
momentum. Conservation of angular momentum prevents
the vortex loops from shrinking and eventual disappearing. This class
of quasiparticles which generally possess nonzero angular momentum 
and charge are called vortons. The presence of the AF condensate is crucial, 
as it is what allows the vortons to carry angular momentum and become
classicaly stable quasiparticles. 

The phenomenon where a condensate forms in the core of a vortex, such that
the vortex can form a spinning loop leading to classical stability, is  
not by any means a new phenomenon. This idea was considered long ago
in the context of cosmology and cosmic strings 
\cite{witten,hindmarsh,turner,davis,davisshell1,davisshell2,
davisshell3,drumvorton,shellard2002,Achucarro:1999it,defectsbook}. 
Our contribution here is the application of the previously developed 
(for cosmic strings and high density QCD) technique to  
the high $T_c$ superconductors.

This paper is organized as follows. In section II we will review the work
of \cite{zhang,arovas,berlinsky,cline} 
where dSC vortices with AF cores  were  originally presented. A comparison
will be made between these vortices and other vortices with 
nontrivial core structure present in high density QCD \cite{superk}.
Although no new results are obtained in this section, we believe
that making a correspondence between two very different fields 
of physics  is quite a useful exercise. In particular, applying the
topological (and some analytical) arguments developed in \cite{superk}
we reproduce the result \cite{zhang,arovas,berlinsky,cline} 
that there is a critical value 
for the coupling constant above which the AF core is not developed. 
In a sense it is a new explanation of the phenomenon based on analytical 
(rather than numerical) calculations.
Section III will contain our new results  where we will show
that classically stable quasiparticles  called vortons are present
within the $SO(5)$ theory of high temperature superconductivity. 
Section IV  will end with concluding remarks and possible experimental 
signatures of the quasiparticles. We also formulate a conjecture that 
the vortons are responsible for the phase transition between AF and dSC
phases at zero external magnetic field when the doping parameter $\mu$ 
is changed.


\section{Vortices with nontrivial core structure- QCD vs. high 
temperature superconductivity}

We will begin by briefly describing the work of \cite{superk} where vortices
with nontrivial core structure were described within the context of QCD
at large baryon chemical potential. This will be beneficial in order to 
make analogies between condensed matter physics and particle physics. We 
should note that this is only a review section and no new material will
be presented here. However, the analogy discussed below will prove to 
be useful for the analysis which follows.
We will then continue 
with a review and comparison  of the vortices with nontrivial core 
structure which 
appear within the $SO(5)$ theory of high \Tc superconductivity 
\cite{zhang,arovas,berlinsky,cline}.
\subsection{Vortices in high density QCD}
There has been a large amount of interest within the 
particle physics community on the subject of QCD, the theory of the strong 
interaction, at large baryon density \cite{colorsc,CFL} (for many 
more references and a nice review see \cite{colorscrev}). At zero baryon
density, QCD is a theory of quarks and gluons which are strongly coupled,
such that confinement takes place and the observable particles are 
colorless hadrons rather than quarks and gluons.
As one increases the baryon chemical potential  the new superconducting
phase when the baryon symmetry is spontaneously broken occurs. 
To explain this phenomenon, let us remind that
in QED, the electron- electron interaction
is in general repulsive, and superconductivity is a very subtle effect. 
In non-Abelian theory, QCD, simple one gluon exchange is always
attractive in the color $\overline{3}$ channel. As is well known from
conventional BCS theory of superconductivity, 
an arbitrarily small attractive interaction will 
lead to the formation of condensate of Cooper pairs near the Fermi surface.  
This is in fact what happens  in QCD at large baryon density. 
The ground state of the high density phase of QCD
is characterized by a diquark condensate \cite{colorsc,CFL} 
analogous to the condensate of electron Cooper pairs 
present in a conventional superconductor. This phase of QCD is 
referred to as a color superconducting phase. The typical chemical potential 
where this phase is thought to occur ($\mu \sim 500$ MeV, 
$\Delta \simeq 100$ MeV, temp $T_c \sim 0.6~ \Delta$, where $\Delta$ is 
superconducting gap) 
cannot be realized
on Earth. The interest in this region of the QCD phase
diagram is motivated by the fact that such densities may be realized
within the core of compact stars, such as neutron stars \cite{NASA}. 

We will not go into specific details, only
state a few of the main features of the color superconducting phase
of high density QCD. 
For the number of quark flavors $N_f=3$ 
(up, down, and strange)
and the number of colors $N_c=3$, the dominant part of the 
diquark condensate takes the following form \cite{CFL}:
\bea 
\label{qqCFL} 
  \<q^{ia}_{L\alpha} q^{jb}_{L\beta} \>^* &\sim&   
  \epsilon_{\alpha\beta\gamma} \epsilon^{ij}\epsilon^{abc} X_c^{\gamma} , 
  \nonumber \\ 
  \<q^{ia}_{R\alpha} q^{jb}_{R\beta} \>^* &\sim&   
  \epsilon_{\alpha\beta\gamma} \epsilon^{ij}\epsilon^{abc}Y_c^{\gamma} , 
\eea 
where $L$ and $R$ represent left and right handed quarks, $\alpha$, $\beta$, 
and $\gamma$ 
are the flavor indices, $i$ and $j$ are spinor indices, $a$, $b$, and $c$ are  
color indices, and $X_c^{\gamma}$ and $Y_c^{\gamma}$ 
are complex color-flavor matrices describing the Goldstone bosons. 
This diquark condensate breaks the original symmetry group
$SU(3)_c \times SU(3)_L \times SU(3)_R \times U(1)_B \times U(1)_A$ (color
gauge symmetry, left and right flavor symmetries, baryonic symmetry, 
and axial $U(1)_A$ symmetry) down to the diagonal subgroup $SU(3)_{c+L(R)}$. 
This diagonal subgroup tells us that whenever we perform an $SU(3)$ color
rotation, we must simultaneously perform and left (right) handed flavor
rotation. Since color rotations are now linked with flavor rotations, 
this phase of high density QCD with $N_f=N_c=3$ is referred to as the 
color-flavor locked phase (CFL) \cite{CFL}. 
Counting the number of broken generators, we see that there
should be 18 Goldstone bosons. Of these 18 GB, 8 of them are eaten by the 
Higgs mechanism resulting in all 8 gluons acquiring a mass. This leaves 
10 GB, an octet related to the breaking of $SU(3)$ and two singlets related
to $U(1)_B$ and $U(1)_A$. 
All of  these bosons (except the one related to $U(1)_B$) are 
actually pseudo-Goldstone bosons due to the small explicit violation
of the symmetry.
In order to describe the low energy degrees of freedom, namely the octet
of Goldstone bosons, one can construct the following gauge invariant field:
\beq 
\label{compsigma}
\Sigma_{\gamma}^{\beta} = \sum_c X_c^{\beta} Y^{c *}_{\gamma} =
	\exp(i \pi^a \lambda^a/ \fpi ),
\eeq 
with the $SU(3)$ generators $\lambda^a$ normalized as 
$\trace(\lambda^a \lambda^b)=2 \delta^{ab}$ and 
$\fpi^2 \sim \mu^2 / (2 \pi^2)$ being the pion decay
constant which can be calculated in the large $\mu$ limit. 
Prior to the work of 
\cite{schafer,bedascha,kaplredd}, it 
was believed that the ground state of the CFL phase was given by 
$\Sigma_o = \diag(1,1,1)$. However, it was noticed that for a physical value of
the strange quark mass $(m_s \gg m_u,m_d)$ this may not necessarily be the 
case. In particular, it was argued in that for $m_s > 60$ MeV 
along with the diquark condensate (\ref{qqCFL} ) a new 
$\kzero$ condensation would occur and that $\Sigma_o  = \diag(1,1,1)$ 
as given above 
would no longer represent the true ground state of 
the CFL phase, but rather, the vacuum expectation value
of the non-diagonal elements of $\sim \<\Sigma^1_2\>$ representing
the $\kzero$ GB would get a  nonzero magnitude \cite{bedascha}. 
Therefore, $\Sigma_o$ would be rotated in some different 
direction in flavor space. 
In the physical 
case where isospin symmetry is not exact (i.e. the up and down 
quarks have different masses, $m_d > m_u$) and we have overall 
electric charge neutrality, $\kzero$ condensation occurs. 
The appropriate expression for $\Sigma_o$ describing the $\kzero$ 
condensed ground state can be parameterized as:
\beq
\Sigma_o= \left( \begin{array}{ccc}
     1 & 0 & 0 \\
     0 & \cos \thetak & \sin \thetak e^{-i\phi} \\
     0 & -\sin \thetak e^{i\phi} & \cos \thetak
 \end{array} \right), 
\eeq
where $\phi$ describes the new Goldstone mode associated with 
$K^0$ condensation along with the diquark condensate discussed above 
and $\thetak \equiv \sqrt{2} \<|K^0|\>/\fpi$
describes the strength of the kaon condensation with \cite{bedascha}:
\bea
\label{vevtheta}
\cos \thetak = \frac{m_0^2}{\mueff^2}, ~~~ \mueff = \frac{m_s^2}{2\mu}, \nn \\
m_0^2 = a m_u (m_d + m_s), ~~~a=\frac{3 \Delta^2}{\pi^2 \fpi^2}.  
\eea
where $m_u,m_d,m_s$ are the masses of the up, down, and strange quarks
respectively, $\mu$ is the chemical potential, and $\Delta$ is the 
size of the color superconducting gap ($\sim 100~$MeV). 
In order for kaon condensation $(\thetak \neq 0)$ to occur, 
we must have $ m_0 < \mueff$. 
This leads to the breaking of the hypercharge \UY symmetry. 
As discussed in \cite{bedascha,superk},
the lightest degrees of freedom in the CFL$+\kzero$ phase are the
$\kzero$ and $\kplus$ mesons. 
The essential physics of these mesons 
can be captured with the following effective Lagrangian:
\bea
\label{LeffKzero}
{\cal L}&=&|\d_0 \Phi|^2 - \vpi^2 |\d_i \Phi|^2 \nn \\
  	&-& \lambda \left(|\Phi|^2 - \frac{\eta^2}{2}\right)^2 -
	\delta m^2 \Phi^{\+} \tau_3 \Phi.
\eea
where $\Phi=(K^+,K^0)$ is a complex doublet describing the $K^0$ and $K^+$ 
mesons, $\tau_3$ is the third Pauli matrix. 
The constants $\fpi$ and $\vpi$
have been calculated in the leading perturbative approximation and 
are given by \cite{ss,bbs}:
\beq
\fpi^2=\frac{21-8 \ln 2}{18} \frac{\mu^2}{2 \pi^2}, ~~
\, \vpi^2=\frac{1}{3}.
\eeq
The remaining parameters in the effective Lagrangian (\ref{LeffKzero}) have 
been obtained from a more complete description of the octet
of Goldstone bosons \cite{bedascha,superk}:
\bea
\delta m^2&=& \frac{a}{2}m_s(m_d-m_u),   \\
\eta^2 &=&\frac{\mueff^2 -m_o^2}{\lambda}. \nn
\eea
In the case that the parameter $\delta m^2$ in (\ref{LeffKzero}) is zero, 
the Lagrangian is invariant under the symmetry group
$SU(2)_I \times U(1)_Y \rightarrow U(1)$ (broken down to $U(1)$). 
From topological arguments we know that such a
Lagrangian does not possess vortices since the vacuum manifold 
is that of 3-sphere and therefore does not have noncontractible loops. 
In the case that $\delta m^2$ is relatively large, 
then the residual symmetry group is
$U(1) \times U(1) \rightarrow U(1)$ 
and the vacuum manifold is that of circle, leading to 
the formation of classically stable global vortex solutions. 
Since $\delta m^2 > 0$ then it is $K^0$ which forms the normal 
global strings with 
$|K^0(r=0)|=0$ and $|K^0(r=\infty)|=\eta / \sqrt{2}$, where the phase varies
from $0$ to $2 \pi$ as one encircles the core of the vortex. 
From these two limiting
cases, it is clear that there should be some intermediate region that somehow
interpolates (as a function of $\delta m^2$) between the two cases. At some
finite magnitude of $\delta m^2$, an instability arises through the 
condensation of $\kplus$-field inside of the core of the vortex. 
As the magnitude of $m_d-m_u$ decreases,  the size of the core 
becomes larger and larger
with nonzero values of both $K^0$ and $K^+$ condensates inside the core.
Finally, at $m_d=m_u$ the core of the string 
(with nonzero condensates $K^0$ and $K^+$)
fills the entire space, in which case the meaning of the string 
is completely lost, and we are left with the situation when $SU(2)$ 
symmetry is exact: no stable strings are possible. As discussed in 
\cite{superk}, the $K^0$ vortex with a $K^+$ condensate on the core 
can be approximately described by the following ansatz:
\bea
\label{Kstringsoln}
K^0 &=& \frac{\eta}{\sqrt{2}} f(r) e^{i \phi}, \\
\label{Kcondsoln}
K^+ &=& \frac{\sigma}{\sqrt{2}} g(r).
\eea
where $\phi$ is the azimuthal angle in cylindrical coordinates, 
$f(r)$ and $g(r)$ are solutions to the classical equations of motion
obeying the boundary condition $f(0)=0,~f(\infty)=1$ and 
$g'(0)=0,~g(\infty)=0, ~g(0)=1, $ and $\sigma$ gives the size of the condensate
at the string core $(r=0)$ determined by the parameters of the Lagrangian. 
The width of the string  when symmetry is restored in the core,
 is given by $1/\kappa$, where 
$\kappa^2 \sim (\mueff^2 -m_o^2 +\delta m^2)$ (the mass scale for $K^0$). 
The width of the $K^+$ condensate when symmetry is not restored
can be estimated  
as $1/ \beta$, where $\beta^2 \sim 2 \delta m^2$ is the mass 
difference between $K^+$ and $K^0$ off of the vortex. 
From these estimates we see that 
as $\delta m^2 \rightarrow 0$,  the width of the vortex core increases
as explained qualitatively above. In order to estimate the critical
point where $K^+$ condensation occurs, we have studied the dynamics
of $K^+$ in  the background of a $K^0$ global vortex solution. This 
is  done by substituting the string solution (\ref{Kstringsoln}) into 
the energy expression derived from the Lagrangian (\ref{LeffKzero}) and 
keeping only terms which are quadratic order in $K^+$. 
The shift in the energy (per unit length) 
in the background of a $K^0$ vortex is 
then given in dimensionless variables as follows \cite{superk}:
\beq
\label{deltaEkp}
\delta E = \frac{\eta^2 \vpi^2}{2} \, 
        \int d^2\rtil\,g(\rtil)[\hat{O}+\epsilon] g(\rtil),
\eeq
where
\bea
\label{Ohat}
\hat{O} &=& -\frac{1}{\rtil} \frac{d}{d \rtil}(\rtil \frac{d}{d\rtil}) 
        - (1-\cos\thetak) (1-f^2(\rtil)) , \\
\label{epsilonK}
\epsilon &=&  \frac{a}{2} \frac{m_s (m_d-m_u)}{\mueff^2}.
\eea
The problem is reduced to the analysis of the 
two-dimensional Schrodinger equation for a particle in an attractive potential 
$V(\rtil)=-(1 - \cos \thetak)(1-f^2(\rtil))$ with $f(\rtil)$ being the solution
of the classical equation of motion for $K^0$ with
the boundary conditions $f(0)=0$ and $f(\infty)=1$.
Such a potential is negative everywhere and approaches zero at infinity.
As is known from standard course in quantum mechanics \cite{landau},
for an arbitrarily  weak potential well there is always a negative 
energy bound state in one and two spatial dimensions; in three
dimensions a negative energy bound state may not exist. For the 
two dimensional case (the relevant problem in our case) the lowest energy 
level of the bound state is always negative and exponentially small for small 
$\lambda'$. One should note that our specific potential 
$V(\rtil)=-(1 - \cos \thetak)(1-f^2(\rtil))$ which enters (\ref{deltaEkp})
is not literally the potential well, however one can always construct the 
potential well $ V' $ such that its absolute value is smaller than  
$ | V(\rtil) | $ everywhere, i.e. $ | V' | < | V(\rtil) | $
for all $ \rtil $. For the potential well $V'$ we know that 
the negative energy bound state always exists; 
when $V'$ is replaced by $V$ it makes the energy eigenvalue even lower.
Therefore, the operator (\ref{Ohat}) has always a negative mode 
irrespective of the local properties of function $f(r)$.
As a consequence, if $\epsilon = 0$ then the string (\ref{Kstringsoln}) 
is an unstable solution of the 
classical equation of motion, the result we expected from the
beginning from topological arguments. The instability manifests itself 
in the form of a negative energy bound state solution of the corresponding 
two-dimensional Schrodinger equation (\ref{deltaEkp}) irrespective of the 
magnitudes of the parameters.
The problem of determining when $\kplus$-condensation occurs is now reduced to 
solving the 2-d Schrodinger type equation $\hat{O} g = \hat{E} g$. 
From the previous discussions we know that $\hat{E}$ for the ground state 
is always negative. However, to insure the  instability  with respect 
to $\kplus$-condensation one should require a relatively large negative 
value i.e.
$\hat{E}+\epsilon < 0 $. It can not happen for arbitrary weak coupling constant
$\sim (1-\cos\thetak)$ when $\thetak $ is small. However, it does happen 
for relatively large $\thetak$. To calculate the minimal critical
value   $\thetacrit$ when $\kplus$-condensation develops, one 
should calculate the  eigenvalue $\hat{E}$ as a function of parameter 
$\thetak$ and solve the equation $\hat{E}(\thetacrit)+\epsilon = 0 $. 
For very small coupling constant 
$\lambda'=(1-\cos\thetak)\rightarrow 0$ the bound state energy is 
negative and exponentially small, $\hat{E}\sim -e^{-\frac{1}{\lambda'}}$.
However, for realistic parameters of $\mu ,~\Delta,~ m_s,~ m_u,~ m_d$ the 
parameter $\epsilon$ is not very small and we expect that
in the region relevant for us the bound state energy $\hat{E}$ is 
the same order of magnitude
as the potential energy $\sim \lambda'$. In this case we estimate  $\thetacrit$
from the  following conditions $-\hat{E}(\thetacrit)\sim\lambda'\sim
(1-\cos\thetacrit)\sim \epsilon$ with the result
which can be parametrically represented as
\beq
\label{K15}
\sin\frac{\thetacrit}{2}
\sim \mathrm{const} \frac{\Delta}{m_s}\sqrt{\frac{(m_d-m_u)}{m_s}},
\eeq
where we have neglected all numerical factors in order to explicitly
demonstrate the dependence of $\thetacrit$ on the external parameters. 
The limit of exact isospin symmetry, which corresponds to
$m_d\rightarrow m_u $ when the string becomes unstable,  
can be easily understood
from the expression (\ref{K15}). Indeed, in the case
that the critical parameter $\thetacrit \rightarrow 0$ becomes 
an arbitrarily small number the $K^+$ instability
would develop for arbitrarily small $\thetak >0 $.

The region occupied by the $K^+$ condensate at this point is determined by the
behavior of lowest energy mode $g$ at large distances, 
$g(\rtil\rightarrow\infty)\sim\exp(-\hat{E}\rtil)$ such that a typical 
$\rtil\sim (m_d-m_u)^{-1}\rightarrow \infty$ as expected. 
\subsection{Vortices in the $SO(5)$ theory of high temperature 
superconductivity}
We will now review the work of
\cite{zhang,arovas,berlinsky,cline} where it was shown that vortices with 
nontrivial core structure similar to the ones discussed above
for high density QCD are present within the $SO(5)$ theory 
of high \Tc superconductivity. 
The  effective Lagrangian which describes
Neel vector $\neel$ and dSC order parameter 
$\psi$ in the presence of zero external electromagnetic field is given 
by \cite{zhang}: 
\bea
\label{L}
{\cal L}  &=&  \frac{\chi}{2}( |\d_t \neel|^2 + |\d_t \psi|^2 )
	- \frac{\rho}{2}(|\nabla \neel|^2 + |\nabla \neel|^2 )\nn \\
   &+&  (\tilde{g} -a) |\neel|^2 -a |\psi|^2 \nn \\
   &-& \frac{1}{2} b |\neel|^4 -\frac{1}{2}b |\psi|^4 -b |\neel|^2|\psi|^2, 
\eea
where we neglected the
electromagnetic contribution to the vortex structure.
Actually, one can show that the electromagnetic field does not change the 
qualitative effects
which are the subject of the present paper, and therefore, will be ignored 
in what follows. Here $\chi$ is the susceptibility and $\rho=\hbar^2 / m^*$ 
is the stiffness parameter. 
In reality, we know that the properties of $\chi$ and $\rho$ 
are different in different directions. 
However, when we discuss the topological 
properties of the configuration this difference can change the
quantitative results but cannot change the qualitative picture. 
The Neel vector has three spatial  components
$\neel = (m_1, m_2, m_3)$ and the superconducting order parameter is 
a complex field $\psi = \psi_1 + i \psi_2$.
The parameters of the above effective Lagrangian are given by:
\bea
\label{Leffparam}
a &<& 0,~~~ b > 0, \\
\label{gtilde}
\tilde{g}&=& 4 \chi (\mu_c^2 - \mu^2).  \\
\label{corrlength}
\xi &=& \sqrt{\frac{\rho}{2 |a|}} 
\eea
where $\mu$ is the chemical potential (or doping, not to be confused 
with the chemical potential for QCD in the previous section),  
$\mu_c$ is the critical
chemical potential which defines the AF-dSC phase boundary, and 
$\xi$ is the coherence length. 
The anisotropy $\gtil$ is included which explicitly breaks the $SO(5)$ symmetry
in the following fashion, $SO(5) \rightarrow SO(3) \times U(1)$.    
If $\gtil = 0$ then the $SO(5)$ symmetry is restored 
and the order parameters $\neel$ and $\psi$ can be organized into a 
superspin order parameter
$\vec{n} = (\psi_1,m_1,m_2,m_3,\psi_2)$ which transforms
in the vector representation of $SO(5)$, as Zhang originally proposed
in \cite{zhang}. 
In the following we will consider $\mu > \mu_c$ and  $\tilde{g} < 0$
so that we are in the dSC phase and 
$\<|\psi|\>=\sqrt{|a|/b} $ and $\<|\neel|\> =0$ in the bulk. 
One immediately notices that the 
form of the Lagrangian for the $SO(5)$ theory is very similar to the 
Lagrangian used to describe $K$ strings in high density QCD (\ref{LeffKzero})
in the previous section. In particular, the key element in construction
of the vortices with non-zero condensate in the core, 
the asymmetry parameter,   is determined by the magnitude of 
$\delta m^2$ in eq. (\ref{LeffKzero}).
For the $SO(5)$ theory it is replaced by the parameter of anisotropy $\gtil$ 
in eq. (\ref{gtilde}).

A nonzero vacuum expectation value for $\psi$ signals the onset 
of superconductivity and the breaking of the $U(1)$ symmetry. It is 
well known that stable vortices can form 
since the topology of the vacuum manifold is that of a circle. 
Analogous to
the situation for high density QCD, for a certain range of the 
anisotropy parameter $\gtil$ these vortices should have an antiferromagnetic
core ($\<|\neel(\rvec=0)|\> \neq 0$). This was initially 
pointed out by Zhang \cite{zhang} when he introduced the $SO(5)$ model
and further studied in \cite{arovas,berlinsky,cline}. Similar to the $K$ 
vortex/condensate solution given by (\ref{Kstringsoln},\ref{Kcondsoln}), 
these vortices are described by the following static field configurations:
\bea
\label{psisoln}
\psi &=& \sqrt{\frac{|a|}{b}} f(r) ~ e^{i \phi}, \\
\label{neelsoln}
\neel &=& \sigma \sqrt{\frac{|a|}{b}} g(r) ~ \hat{m},
\eea
where $\phi$ is the azimuthal angle in cylindrical coordinates, 
$\sigma$ is the parameter obeying the relation $0 \leq \sigma < 1$, 
and $\hat{m}$ is an arbitrary unit vector. As before, $f(r)$ and 
$g(r)$ are solutions to the classical equations of motion satisfying
the boundary conditions $f(0)=0,f(\infty)=1$ and 
$g'(0)=0,~g(\infty)=0,~g(0)=1$. 
The width of the vortex determined by the profile function 
$f$ is approximately given by the 
coherence length (\ref{corrlength}), 
$\delta_{\psi} \approx \xi$. The width of the 
condensate in the core (if it forms)
is estimated to be of order of 
$\delta_m \approx 1/\sqrt{|\gtil|}\sim 1/\sqrt{(\mu^2 - \mu_c^2)}$
and becomes very large at the phase boundary.

Using what we have already learned from QCD and the results from 
earlier work on these vortices \cite{arovas,berlinsky,cline}, 
we can immediately
summarize the main features of these objects. Numerical calculations 
in \cite{berlinsky} confirm that as the anisotropy parameter $|\gtil|$ 
is decreased, the size and width of the condensate in the  core increases. 
We will support these numerical calculations using some analytical 
arguments given below. 

The free energy (per unit length) 
obtained from the Lagrangian (\ref{L}) is:
\bea
\label{freeenergy}
\frac{{\cal F}}{l} &=& \int d^2 r 
	\left[  \frac{\chi}{2}\left( |\d_t \neel|^2 + |\d_t \psi|^2 
\right)  \right.
	+ \frac{\rho}{2}(|\nabla \neel|^2  \nn \\
   &+& |\nabla \psi|^2 ) - (|a|+\gtil) |\neel|^2 -|a| |\psi|^2 \nn \\
   &+& \frac{1}{2} b |\neel|^4 
	\left. +\frac{1}{2}b |\psi|^4 +b |\neel|^2|\psi|^2 \right].
\eea
If anisotropy $\tilde{g}\equiv 0$ we know from topological arguments  that this
theory does not possess any vortices since the vacuum manifold 
is that of a 4-sphere and therefore does not have noncontractible loops. 
If  $|\tilde{g}|$ is relatively large, 
then the residual symmetry group has a subgroup $U(1)$ 
and the vacuum manifold is that of circle, leading to 
the formation of classically stable global vortex solution described 
in terms of $\psi$ field
 with a typical profile function
 when 
$|\psi(r=0)|=0$ and $|\psi(r=\infty)|= \sqrt{|a|/b}$. 
From these two limiting
cases, as discussed in the previous subsection
IIA,  it is clear that there should be some intermediate region that somehow
interpolates (as a function of $\tilde{g}$) between the two cases.
The way how this interpolation works is as follows
(see the previous subsection where the physical
picture is quite analogous to the present case).
 At some
finite magnitude of $\tilde{g}$, an instability arises through the 
condensation of $\neel $-field inside of the core of the vortex. 
As the magnitude of $\tilde{g}$ decreases,  the size of the core 
becomes larger and larger
with nonzero values of both $\neel$ and $\psi$ condensates inside the core.
Finally, at $\tilde{g}=0$ the core of the string 
(with nonzero condensates  $\neel$ and $\psi$)
fills the entire space, in which case the meaning of the string 
is completely lost, and we are left with the situation when the $SO(5)$ 
symmetry is exact: no stable strings are possible. 

In order to estimate that critical value of the parameter 
$\gtil=\gcrit$ where
an AF core forms inside the vortex, the same method can be applied 
as described for the QCD color superconductor
in the previous section. We will use the following
change of variables in order to express the free energy in terms of
dimensionless variables only:
\bea
\label{dimvar}
\psi = \sqrt{\frac{|a|}{b}} \psi',~~ 
\neel = \sqrt{\frac{|a|}{b}} \neel',~~ \vec{r} &=& \xi \vec{r}'.
\eea
Expanding the expression for the change in the free energy 
(\ref{freeenergy}) in the background of a $\psi$ vortex solution
given by (\ref{psisoln}) and keeping only quadratic terms in 
$\neel$, we have:
\beq
\label{deltaF}
\frac{\delta {\cal F}}{l} 
	= \sigma^2 \frac{\rho |a|}{2 b}
        \int d^2 r' \,g(r')[\hat{H}+\epsilon] g(r'),
\eeq
where
\bea
\hat{H} &=&  -\frac{1}{r'} \frac{d}{dr'}(r' \frac{d}{dr'}) -(1- f^2(r')), \\
\epsilon &=&  \frac{|\gtil|}{|a|} = 4 \frac{\chi (\mu^2 - \mu_c^2)}{|a|}.
\eea
Since we are working in the dSC phase $\gtil < 0$  the 
perturbation $\epsilon > 0$. We have now cast the change in the 
free energy in the exact same form as we did for the QCD vortices in 
the previous section. The problem is now reduced to the analysis
of the two-dimensional Schrodinger equation for a particle in an 
attractive potential $V(r')= -(1- f^2(r'))$. As before, this 
potential is negative everywhere and approaches zero at infinity. 
This means that the ground state eigenfunction $\hat{H} g_0 =\hat{E} g_0$
has a negative eigenvalue $\hat{E}$. The instability with respect to formation
of the AF condensate in the core occurs not for arbitrary small negative
eigenvalue $\hat{E}$, but when the absolute value of  $|\hat{E}|$
is large enough to overcome the positive contribution  due to $\epsilon$. 
Therefore, we immediately see that an AF core forms 
if $\hat{E} + \epsilon  \leq 0$.
If $|\gtil|$ is greater than some critical value $\gcrit$ then 
it is not energetically favorable for an antiferromagnetic core to 
form and dSC vortices will possess a normal core where the 
symmetry is restored. 
Following the same procedure as in the QCD case, we have:
\beq
 \frac{|\gtil_{crit}|}{|a|} 
	= 4 \frac{\chi (\mu^2 - \mu_c^2)}{|a|} \simeq 0.2 . 
\eeq 
where for numerical estimates we used the variational 
approach developed in \cite{turner}.

Above we have reviewed the basic properties of superconducting vortices
with an antiferromagnetic core within the $SO(5)$ theory of superconductivity.
We should emphasize once more that all results presented   above 
are not new and have been discussed previously from a different perspective.
Let us repeat the main results of this section once again:
If $\gtil =0$ then the dSC vortices
are unstable. 
If $0 < |\gtil|< \gcrit$ then an AF core will
form inside the dSC vortices. The width of the AF core in this case becomes
larger and larger when we approach the phase transition
line, i.e. $\gtil \rightarrow 0$. 
Finally, if $|\gtil| > \gcrit$ then the 
dSC vortices will have a normal core when symmetry is restored and   $|\neel|(r=0) = |\psi|(r=0) =0$. 
In what follows
we  will always be working in the region of the phase 
diagram where $0 < |\gtil|< \gcrit$ and dSC vortices have an AF core
(which will be referred to as dSC/AF vortices). 
Now we will proceed to the next section and introduce vortons, loops 
of dSC/AF vortices which are stabilized by angular momentum. 


\section{Vortons in high temperature superconductivity}

We will now consider the interesting possibility that loops of the 
dSC/AF vortices can exist as classically stable objects  
( and at least metastable quantum mechanically). 
This stability arises  through
a mechanism where topological and Noether charges can be trapped 
on the core of the vortex. Such objects, called vortons, have been 
studied extensively in the context of cosmology where cosmic strings 
have nontrivial core structure. 
\cite{witten,hindmarsh,turner,davis,davisshell1,davisshell2,davisshell3,drumvorton}. 
Such vortons are also present within 
high density QCD where vortices with a condensate trapped on the 
core are realized \cite{qcdvorton}. 

As Davis and Shellard originally pointed out in 
\cite{davis,davisshell2,davisshell3}, if one has a theory 
which  contains vortices with a condensate trapped on the  core then 
loops of such vortices can form which are stabilized by angular 
momentum alone.  
We will consider a large loop of string of radius $R \gg \delta$, where
$\delta$ is the vortex thickness, so that curvature effects can be
neglected. The $z$-axis is defined along the length of the string,
varying from $0$ to $L=2 \pi R$ as one goes around the loop. Although
we are considering a circular loop for  simplicity at the moment, we 
realize that this is probably not the relevant physical case. 
The results we will 
discuss in this section should not depend on the geometry of the loop, the
important point is the presence of conserved charges which are trapped 
on the vortex leading to stability. In reality, the final stable configuration
of these vortex loops is probably a more complicated shape because of the 
quasi two-dimensional nature of the high temperature superconductors. In 
particular, we have neglected the difference between the transverse and 
tangential spatial directions in our treatment of the problem. 
The appropriate calculations would include this difference and lead 
to a non-symmetric shape. However, we neglect these complications at 
this stage. 

In order to make an analogy with the QCD  case where the condensate 
on the core is described by a complex field, we are free to represent 
two degrees of freedom represented by a unit
Neel vector $ \hat{m}, ~\hat{m}^2=1 $ defined by eq. (\ref{neelsoln})
 in terms of a single complex field  $\Phi$ as, 
\beq
\label{mhat}
\hat{m} = \left( \frac{\Phi + \Phi^*}{1 +|\Phi|^2}, 
	\frac{\Phi -\Phi^*}{i(1 +|\Phi|^2)}, 
	\frac{1 -|\Phi|^2}{1 +|\Phi|^2} \right),
\eeq
where $\Phi = |\Phi| e^{i \alpha}$ (this is simply the projection of the 
unit sphere onto the complex plane).  
At this point we are  free to pick the direction of the Neel vector. 
For a background classical  field describing a 
vortex defined along the $z$ direction, we will
pick $\neel$ to lie in the $xy$-plane so that $m_z=0$.
We should note that all calculations and results which 
follow do not depend on the particular choice of $\neel$ that we have made 
above. However, we do expect that this will turn out to be the lowest energy 
configuration when higher order derivative terms are included 
in the free energy (see below). 
We neglect  fluctuations of the absolute value $|\Phi|$ 
for description of the classical background and consider
variation of its phase $\alpha$.
In this case, we have $|\Phi| = 1$ for
the classical background field 
as it follows from the transformation (\ref{mhat}), and 
$\neel$ simplifies to:
\beq
\label{neelxy}
\hat{m} = \frac{1}{2} \left( \Phi + \Phi^*, 
	i(\Phi^* -\Phi), 0 \right).
\eeq
with $|\Phi| =1$ to be fixed.
The condensate $|\Phi| \neq 0 $ can carry currents and charges 
along the string so we will represent it by the following ansatz 
which describes the dependence of these excitations on $z,t$:
\beq
\label{Phi}
\Phi = |\Phi|e^{i \alpha(z,t)} = |\Phi|e^{i (k z - \omega t)}. 
\eeq
With this redefinition of the fields, the kinetic term esquires two
additional terms due to the $(z, t)$ dependence of the phase
in the core:
\bea
\label{newkin}
\frac{\chi}{2}\left[ |\d_0 \neel|^2 - v_s^2 |\d_z \neel|^2
\right] \rightarrow \\ \nn
	  \frac{\chi}{2}m^2 (r) \left[(\d_0 \alpha(z,t))^2 
	-v_s^2  (\d_z \alpha(z,t))^2\right], 
\eea
where $v_s^2 \equiv \rho / \chi$.
The key point of the time dependent ansatz (\ref{Phi}) is as follows.
Naively, one could think that the time dependence in  a classical
solution brings an additional energy into the system
which usually does not help to stabilize
the configuration. However, as Witten noticed in \cite{witten}
if there is a conserved charge in the system, the configuration
could be stable due to the conservation of the corresponding charge.
In a sense, the time-dependent configuration becomes the lowest
energy state in the sector with a given non-zero charge. A similar 
time-dependent ansatz for a different problem 
was also discussed by Coleman in 
\cite{qball} where he introduced so-called Q-balls, macroscopically large
stable objects with a time dependent wave function.
We follow ref. \cite{witten} and  define a charge 
$N$ which is topologically conserved:
\beq
\label{topological}
N = \oint_C \frac{dz}{2 \pi} \left(\frac{d \alpha}{d z}\right) = k R,
\eeq
where the path $C$ is defined along the vortex loop and we assume 
that $\omega,k$ are some constants along the loop. 
Since $\alpha$ can change by multiple of $2 \pi$ in circling the 
vortex loop, $N$ must be an integer. This is required in order for 
the condensate $m$ to remain single valued. 

In addition to the topologically conserved winding number $N$, 
there also exist the standard Noether charges and currents which  can be 
trapped on the vortex core associated with the parameter $\omega,k$ included
in the phase $\alpha$ above. 
In our case, the relevant symmetry, $SO(3)$, implies conservation
of three Noether charges:
\bea
\label{noether}
Q_k = \int d^3 r j^0_k 
    = i \chi \int d^3 r [ (\d_0 m_a) ~(S_k)_{ab}  ~m_b ],	
\eea
while the corresponding three currents are:
\bea
\label{current}
J^z_k =\int d^3 r j^z_k ,  
    = -i \rho \int d^3 r [ (\d_z m_a) ~(S_k)_{ab}  ~m_b ],
\eea
where $S_k$ are the three generators of $SO(3)$. 

A vortex loop with nonzero Noether charges $Q_k$ and topological charge
$N$ trapped on the core is 
described by an ansatz of the following form which depends 
on the position $\rvec=(r,\phi,z)$ and time $t$ as
(using Eq.'s (\ref{psisoln}),(\ref{neelsoln}),(\ref{mhat}), and (\ref{Phi})):
\bea
\label{psivorton}
\psi = \sqrt{\frac{|a|}{b}} f(r) ~ e^{i \phi}, 
\eea
\bea
\label{mvorton}
\neel = \sigma \sqrt{\frac{|a|}{b}} g(r) 
	\left(\cos (k z - \omega t),~  
	  \sin (k z -\omega t),~0\right),
\eea
where as in the previous section $f(r),g(r)$ are solutions to the classical
equations of motion obeying the appropriate boundary conditions. 

For the solution given by (\ref{mvorton}) we have a 
nonzero  Noether charge $Q_z$ which is trapped on the vortex core:
\beq
\label{Qz}
Q_z = \chi L \omega \Sigma, 
\eeq
where $\Sigma$ is defined as as the integral of 
$|\neel|^2$ over the vortex cross section:
\beq
\label{Sigma}
\Sigma = \int_\times d^2 r ~|\neel|^2. 
\eeq
We should note that for a different choice of the direction of Neel vector
$\hat{m}$.  
the conserved charge which is nonzero would be different. 
The important point is that a nonzero charge will always be present 
independent of the of the Neel direction $\hat{m}$. 

As we mentioned at the beginning of this section, these vortons
are spinning and carry a non-zero angular momentum. The vortons are 
stable  against shrinking due to conservation of 
angular momentum. 
To calculate the angular momentum of a vorton with nonzero 
charges $N,Q_z$ trapped on the core, we use the standard formula
for the angular momentum expressed in terms of the energy-momentum tensor:
\beq
\label{energymomtensor}
M_{ij} = \int d^3 r (T_{0i} x_j - T_{0j} x_i),
\eeq
which can be approximated for a large vorton in the plane as:
\bea
\label{angmom}
M &\simeq& 2 \pi \chi R^2 \sigma^2 \frac{|a|}{b} 
	\int d^2 r g(r)^2 \omega k , \nn \\
  &=& 2 \pi \chi R^2 \omega ~k ~\Sigma. 
\eea
The angular momentum points
in the direction normal to the surface formed by the vorton. We now see 
that the reason we would expect such configurations to be 
classically stable is simple, it is just because these
vortons are spinning and angular momentum is conserved.
One can say that the vorton is stable because it is the lowest
energy configuration in the given sector with nonzero conserved charges
$N,Q_z$.   Angular momentum $M$ (\ref{angmom}), which is essentially the 
product of two charges $N$ and $Q_z$ is also nonzero when both
charges  $N$ and $Q_z$ are non-zero. In the discussion above 
we neglected the higher order derivative terms. In particular, there 
will be some correlation between the charge $Q_i$ (\ref{noether}) 
and the momentum along the vorton  $P_i= T_{0i}$
in the expression for the free energy ($ \sim T_{0i} Q_i$). Such 
a correlation implies that the ansatz (\ref{neelxy}) will represent the lowest
energy configuration if the angular momentum $\vec{M}$ (\ref{angmom})
points in the direction normal to the surface formed by the vorton. 

We will assume that we now have a vorton configuration with nonzero 
values of  $N$ and $Q_z$. In order to assign specific numbers for these
quantities one must look at the mechanism of formation. We will not 
address such complex issues in this paper and simply assume that there
is some nonzero probability  for  a vorton to form. For recent work 
on the issue of vorton formation we refer the reader to 
\cite{formation}. The free energy of a vorton can be obtained by
substituting (\ref{psivorton}) and (\ref{mvorton}) into 
(\ref{freeenergy}):
\bea
\label{freeenergyvorton}
{\cal F} &=& \int d^3 r \left[\frac{\chi}{2}  \right.
	((\omega^2 + v_s^2 n^2) m^2  + v_s^2(\nabla_r m)^2 \nn \\
   &+& v_s^2 |\nabla \psi|^2)) - (|a|+\gtil) m^2  - |a| |\psi|^2 \nn \\
   &+& \left. \frac{1}{2} b (m^4 
	+|\psi|^4 +2 m^2|\psi|^2) \right],
\eea
where $m = \sigma \sqrt{|a|/b} ~g(r)$. 
We can simplify  this expression further by using the fact that 
$m$ is a solution to the equation of  motion and represent the free energy
in the following way:
\beq
\label{freesimple}
{\cal F} = L \left(  \rho \pi \frac{|a|}{b} \ln (\Lambda/\xi)
	-\frac{b}{2} \Sigma_4 + \chi  \omega^2 \Sigma \right) ,
\eeq
where  we have  defined the quantity $\Sigma_4$ for brevity:
\beq
\Sigma_4 = \int_\times d^2 r |\neel|^4. 
\eeq
The first term in (\ref{freesimple}) is simply the energy from the dSC 
vortex with no condensate present in the core (to logarithmic accuracy). 
Here $\Lambda$ is the long distance cutoff which must be included
to regulate the logarithmic divergence of the normal global string. 
The long distance cutoff is typically the distance between vortices, 
so in our case we will take $\Lambda = L$ where $L$ is the length 
the vortex loop. The 
second term is negative, reflecting the fact that it is energetically 
favorable to have an AF core. And the  third term is the additional 
contribution to the energy due to nonzero $Q_z~(N)$. 

There are various cases which must be considered,
 $v_s k > \omega, v_s k < \omega$, and $ v_s k = \omega$. 
Notice that the effect of adding 
a having $k, \omega$ nonzero is the addition of a masslike term 
for $\neel$ to the Lagrangian:
\beq
\label{deltaL}
\delta {\cal L} = \frac{\chi}{2 }(\omega^2 - v_s^2 k^2) m^2. 
\eeq
If $v_s k > \omega$ then the effect of 
a nonzero $v_s k,\omega$ is to add a positive mass term for 
$|m|$ to the Lagrangian. 
This counteracts the effects of the 
negative mass term in the original Lagrangian (\ref{L}). 
Since $k \sim 1/L$ quenching occurs and 
the size of the condensate $\Sigma$ 
decreases as the vortex loop gets smaller. Conversely, if $\omega > v_s k$
one has the opposite situation and anti-quenching occurs. As the 
vortex loop shrinks, the size of the condensate $\Sigma$ gets larger
as one would expect. The different cases
have been examined using numerical calculations in \cite{davis, davisshell2}. 
Recall (\ref{Qz}) that $\omega$ is given as
\beq
\omega = \frac{Q_z}{\chi L} \Sigma^{-1}.
\eeq
As Davis and Shellard point out in \cite{davisshell2}, if $\omega$ starts
out less than $v_s k$ quenching occurs and forces $\omega$ increase faster than
$Q_z/ L$. In the opposite case where $\omega > v_s k$ anti-quenching  occurs
and therefore $\omega$ increases more slowly than $Q_z/ L$. 
The important conclusion that was drawn from this analysis
is that $\omega/(v_s k) \rightarrow 1$ is an attractor 
\cite{davisshell2}. As a loop shrinks $\omega/ (v_s k)$ approaches $1$ and the 
quenching (or anti-quenching) slows and the eventually stops leaving a 
classically stable vorton behind. 
Therefore, for simplicity 
we will focus on the so called
chiral case when $\omega = v_s k$ which is the most stable configuration.
Realistically, we know that the periodic structure of 
the material of the superconductor breaks the rotational symmetry, 
leading to non-conservation of angular momentum (it can be transferred
to the material). However, the topological
charge $N$ is still a conserved quantity. Therefore, for the chiral
vortons, $Q \sim N$ is also conserved due to the relations 
(\ref{topological}) and (\ref{Qz}). For such configurations, stability
is ensured. When $\omega \neq v_s k$, 
the transfer of angular momentum to the lattice is possible, eventually 
settling to the chiral case with $\omega = v_s k$. 

In the chiral case the 
size of the condensate $\Sigma$ is independent of $N,Q_z,L$ and 
the free energy (\ref{freesimple}) can be written as:
\bea
\label{tension}
{\cal F} &=& L \alpha_s  + (2 \pi)^2 \rho \frac{N^2 \Sigma}{L}, \\
\alpha_s &=& \rho \pi \frac{|a|}{b} \ln (\Lambda/\xi) 
	-\frac{b}{2} \Sigma_4, \nn
\eea
where $\alpha_s$ is the string tension of the bare dSC/AF vortex
with $Q_z=N=0$ and $k$ is expressed in terms of the conserved 
winding number $N$ according to eq. (\ref{topological}).
Written in this form, it is immediately obvious that for a given nonzero 
value of $N$ the free energy has a minimum at $L = L_{0}$:
\bea
\label{Lmin}
\frac{N}{L_0} 
	= \frac{1}{2\pi}\sqrt{ \frac{\alpha_s}{\rho\Sigma} },
\eea
We can give   a crude estimate of the winding number density,
$n_0\equiv \frac{N}{L_0} $,  of a stable vorton configuration:
\bea
\Sigma &\sim& \sigma^2 \frac{|a|}{b} \delta_m^2, \\
\alpha_s &\sim& \rho \frac{|a|}{b}, 
\eea
where $\delta_m$ is the width of the condensate. This gives us:
\beq
\label{n_0}
n_0\equiv \frac{N}{L_0}  
	\sim \frac{1}{\delta_m}\sim \sqrt{ |\tilde{g}|}
\sim\sqrt{ { 4 \chi (\mu^2 - \mu_c^2)}}, 
\eeq
which is approximately the inverse width of the condensate. 
As expected, the winding number 
density $n_0 $ does not depend on the large number $N$, 
but depends only on the internal structure
of the vorton, i.e. on the width of AF condensate in dSC vortex core. 
The equation (\ref{n_0})  tells us that as one goes 
around a vorton the direction of 
the Neel vector $\hat{m}$ varies over a distance scale $\sim \delta_m$, 
the width of the condensate. As the 
doping is decreased and the AF-dSC phase boundary is 
approached from above the width of the condensate increases. For a 
given value of $N (Q_z)$ (determined at the time of formation) the 
size of a stable vorton increases as one approaches the AF-dSC phase 
boundary. 

The discussion above has shown that vortons are indeed classically stable. 
This would imply that on the quantum mechanical level such quasiparticles
are at least metastable. The issue of quantum stability of vortons 
was addressed in a recent paper \cite{vilenkin}. In this paper  they 
calculated the lifetime of a vorton with $Q=0,N \neq 0$ (pure current case). 
The mechanism of decay is some quantum mechanical tunneling process 
where the condensate instantaneously 
goes to zero on the core, allowing the winding
number to decrease by one unit from $N$ to $N-1$. 


\section{Conclusion and Further Speculations} 

In this paper we have reviewed the dSC vortices which have an 
antiferromagnetic core within the $SO(5)$ theory of high 
temperature superconductivity \cite{zhang,arovas,berlinsky,cline}. 
We have compared these dSC/AF vortices with similar vortices which 
arise in a completely different context, high density QCD 
\cite{fzstrings,krstrings,superk}. 

The main point that was presented in this paper is that loops
of dSC/AF vortices called vortons can exist as classically stable objects in 
the presence of zero external magnetic field. 
The source of the stability of these vortons
is conservation of angular momentum that 
counteracts the string tension, which prefers to minimize the 
length of the vortex loop. The fact that there is a condensate trapped
on the vortex core is crucial for the stability of vortons. It is 
the condensate which allows nonzero charges to be trapped on the core, 
leading to the presence of nonzero angular  momentum. It remains to 
be seen if such quasiparticles  will be important for the physics of 
the high \Tc superconductivity. In what follows, we provide 
arguments supporting the idea that the vortons can play a  
key role in AF-dSC phase transition. At this point we consider  
the vorton mechanism driving AF-dSC phase transition as a conjecture.

The first argument goes as follows.
As we have shown above, for a given value of $N$ there exists classically 
stable vorton configurations with size $L$ and fixed 
ratio $N/ L_{0} \sim 1/ \delta_m$, 
where $\delta_m$ is the width of the condensate that is trapped
on the vortex core. As one decreases the doping parameter and 
approaches the AF-dSC phase boundary, the width of the 
condensate $\delta_m$ increases. 
This is the direct consequence of the fact that the 
asymmetry parameter $|\gtil|$
becomes smaller and smaller when the phase boundary is approached.
From the relation $  L_0/N  \sim  \delta_m $ given above, this
would imply that $L_{0}$, the length of a classically 
stable vorton, must increase. Decreasing
the asymmetry parameter $|\gtil|\sim 4 \chi (\mu^2 - \mu_c^2)$ 
further would result in a 
large vorton with large core size. The  volume of the 
regions filled with the AF state
behaves like $L_0  \delta_m^2\sim |\gtil|^{-3/2}$. When the 
phase transition line is approached,
the regions with the AF state fill the entire space. 
At this point the AF-dSC phase transition occurs.
Although our approximations are no longer valid at this point, 
because our description assumes that the vorton core size is 
much smaller than $L$
and the interaction between strings  can be neglected
(plus many other assumptions not to be mentioned). These 
assumptions certainly fail
in the vicinity of the phase transition. Nevertheless, 
the fact that the size of the 
AF regions inside of dSC phase increases rapidly when the 
AF-dSC phase transition is approached 
should be considered as a strong argument in favor of the  vorton mechanism 
driving the AF-dSC phase transition.  

We would also like to make the observation that on the other side of the
phase transition boundary, in the AF state, 
there are quasiparticles whose cores are in dSC phase \cite{sheehy}. 
Therefore, one can imagine a situation when one type of
quasiparticles (dSC vortices with an AF core) becomes a different type of 
quasiparticles (AF skyrmions with a dSC core) when the doping 
parameter decreases and the phase transition line is crossed.

The next natural question to ask is as follows: let us assume that vortons
are indeed the relevant quasiparticles which drive AF-dSC phase transition at 
small temperatures. Can the same vortons  be an essential part of the 
dynamics when the temperature (rather than the chemical potential $\mu$) 
crosses the superconducting phase transition at $T_c$? 
If the answer is positive, we would have a nice unified  
picture for two different phase transitions on the $(T, \mu)$ plane.
We believe the answer, indeed, could be positive (see arguments below).

We start by reminding the reader that the pseudogap phase is characterized  by
the temperature $T_c < T < T^*$, when  
the Cooper pairs are already formed but the long-range phase 
coherence sets in only at
the much lower temperature $T_c \ll T^*$. It is believed that 
in this regime the phase order is destroyed
by  fluctuating vortices of the Cooper pair field $\psi$ above 
$T_c$ \cite{vortices}.
It is quite natural to {\it identify } our vortons (loops of vortices)
sliced by a two dimensional plane  with vortex-antivortex
pairs with distinct experimental signatures
from ref. \cite{vortices}. In this case, since underdoped cuprates are 
effectively two-dimensional,
at finite temperature the loss of phase order may be expected to proceed via
the Berezinsky-Kosterlitz-Thouless phase transition. In this case, 
the vortons discussed 
in the present paper, being sliced by the  two-dimensional plane, become
the vortex-antivortex pairs analyzed in ref. \cite{vortices} and  
could be  responsible 
for the phase transition at $T_c$. However, the picture of the 
phase transition here
is quite different from what we previously discussed regarding 
the AF-dSC  separating line.
In the present case, when $T$ crosses $T_c$ the transition 
happens because of the vortex-antivortex interaction
which is proportional to  $\rho\pi\frac{|a|}{b}\ln(x_1-x_2)$ 
and not because the seeds  of a new phase (the vorton cores) fill the 
entire space. This is the typical two dimensional form due to the global
nature of the vortices (local strings do not
possess this feature).  The volume occupied by the vortex cores
at this point is still much smaller than the volume of the system. 
It is well known that such a 
logarithmic interaction is a key element for understanding
the Berezinsky-Kosterlitz-Thouless phase transition.

Encouraged by the argument given above, we extend our 
conjecture and  assume   that the 
same quasiparticles, vortons, are  
responsible not only for the AF-dSC phase transition 
but also for the phase transition
 separating the pseudogap   and dSC phases at temperature 
$T_c$ when $\mu > \mu_c$.
The natural question to ask is: how does the critical 
temperature $T_c(\mu)$ depend on the
chemical potential $\mu$ within this conjecture? To answer this question we 
recall that the critical temperature $T_c$  for the 
Berezinsky-Kosterlitz-Thouless phase transition is proportional 
to the strength of the logarithmic interaction mentioned above. 
In our case it is nothing
but the string tension determined (mainly) by the vacuum expectation value 
of $\psi$ field, (\ref{tension}), i.e. $T_c \sim \alpha_s$.
When $\mu=\mu_c$ the asymmetry parameter is zero and 
$\<\psi^2\>=|a|/b$ is determined 
by the unperturbed coefficients $a, b$. The simplest way 
to determine how the expectation value of the field 
$\psi$ varies when $\mu$ increases is to introduce
the asymmetry parameter in the ``symmetric" 
manner \cite{foot1}.
${\cal{F}}_{asym}=-\tilde{g}(\vec{m}^2-|\psi|^2)$
such that negative $\tilde{g}$ corresponds to the condensation
of the $|\psi|^2$ field, 
and positive $\tilde{g}$ corresponds to the condensation
of the $|\vec{m}|^2$ field. This asymmetry parameter enters the string 
tension in dSC phase 
as follows, $\alpha_s\sim (|a|-\tilde{g})/b, ~\tilde{g}< 0 $. 
From this expression one can immediately deduce
that $\alpha_s$ (and therefore, $T_c$) will increase with 
$\mu$ when $\tilde{g}$
being negative, becomes larger and larger in magnitude. 
It is interesting to note that
the very same conclusion has been reached in the original 
paper \cite{zhang}, where 
the increasing of $T_c$  with chemical potential was explained as a result
of mass increase of the $\pi$ triplet (see \cite{zhang} for details).


Having presented our arguments supporting the conjecture that 
vortons might be the relevant degrees of freedom in the 
dSC phase at zero external magnetic field, we  
now conclude with a few remarks on 
how this picture can be experimentally tested.
First, the possible experimetal methods  
(such as $\mu SR$ and inelastic neutron scattering)
for observing AF vortex cores 
were discussed in ref. \cite{arovas}  and we shall not repeat their analysis.
However, we should mention that the fact that the AF cores do appear
in the vortices \cite{vaknin,lake,dai,mitrovic,miller}
suggests that the $SO(5)$ model of high $T_c$ 
superconductivity may be correct. 
Our original remark here is that AF vortex core size is on the order of dSC 
coherence length far away from the point $\tilde{g}=0$, and becomes larger
when the chemical potential approaches $\mu_c$ at zero external magnetic 
field. 
We expect that 
the average size $\<L\>$ of vortons
grows with temperature, and therefore, correlations between AF cores
will grow with temperature as well. 
Similar behavior 
is expected to occur when $\mu$ approaches $\mu_c$ at a fixed temperature.
In this case the effect is expected to be even more pronounced
because the volume occupied by the AF cores grows as $|\tilde{g}|^{-3/2}$
as discussed above,
and therefore the correlations as well as the magnitude of the 
local electron magnetic fields should scale accordingly.
This picture suggests that the AF correlation length is propotional 
to $L$ and could be very large, much larger than any other scale 
of the problem. Apparently, such large AF correlation lengths have 
already been observed in \cite{lake} and we would like to argue 
that this correlation is related to our vortons. One should remark 
here that such a large correlation length cannot be simply explained by
the interaction between vortices (which have size $\xi \sim 20~ \AA$) 
because it would lead to a strong dependence on the Neel temperature 
as a function of the intervortex spacing controlled by the external 
magnetic field, while observations suggest that the Neel temperature 
is field independent \cite{lake}. 

All the effects previously mentioned require the existence of the AF 
core in the vortex and they
are not specifically sensitive to the existence of vortons, 
which is the subject of this work.  
The main feature of the vortons is that they can carry angular momentum 
(see eq. (\ref{angmom})) and provide large AF 
correlation lengths (see discussion above). 
Therefore, these excitations should be present 
if a dSC sample is rotated with nonzero angular momentum. 
The situation is very similar to the 
$^3He$ and $^4 He$ systems where superfluid vortices 
can be studied by rotating liquid helium in a can.
In many respects our vortons are similar to rotons, 
and presumably can be studied
in a similar way using the technique developed for these systems.
 
Finally, if vortons are indeed the relevant degrees of 
freedom in the high $T_c$ superconductors, 
it provides a unique opportunity to study cosmology and 
astrophysics by doing laboratory experiments
in condensed matter physics. Over the last few years 
several experiments have been done to test ideas drawn from cosmology 
(see the review papers \cite{kibble,volovik} for further details).

\section*{Acknowledgments}
We are thankful to S. C. Zhang and John Berlinsky for the nice lectures  on 
the $SO(5)$ Model of High 
$T_c$ Superconductivity given at UBC, which  motivated the present study.
We are grateful to Igor Herbut and Dan Sheehy for fruitful conversations.
This work was supported in part by the Natural Sciences and Engineering 
Research Council of Canada and by the European Science Foundation 
Programme ``Cosmology in the Laboratory''.


\begin{thebibliography}{50}
\bibitem{foot1}
This definition of asymmetry is different from what
we used in the rest of the paper.
\end{thebibliography}
\end{document}